\documentclass[fleqn,12pt]{wlscirep}
\usepackage[utf8]{inputenc}
\usepackage[T1]{fontenc}
\usepackage{setspace}
\usepackage{bm}
\usepackage{enumitem}
\usepackage{color}
\usepackage{soul}
\usepackage{lineno}

\title{Five-stage ordering to a topological-defect-mediated ground state in a buckyball artificial spin ice }

\author[1,2,*]{Gavin M. Macauley}

\author[1,2]{Luca Berchialla}

\author[2,3, *]{Peter M. Derlet}%

\author[1,2,*]{Laura J. Heyderman}%

\affil[1]{Laboratory for Mesoscopic Systems, Department of Materials, ETH Zurich, 8093 Zurich, Switzerland}
\affil[2]{Paul Scherrer Institute, Center for Neutron and Muon Sciences, Forschungsstrasse 111, 5232 Villigen, Switzerland}
\affil[3]{Paul Scherrer Institute, Center for Scientific Computing, Theory and Data, Forschungsstrasse 111, 5232 Villigen, Switzerland}
\affil[*]{Corresponding authors: gavin.macauley@psi.ch, peter.derlet@psi.ch,    laura.heyderman@psi.ch}

\begin{abstract}
Artificial spin ices are arrays of coupled nanomagnets, which exhibit a variety of fascinating collective behaviour including emergent magnetic monopoles, charge screening, and novel phase transitions.
However, they have mainly been confined to two dimensions due to the challenges inherent to their fabrication and characterisation in three dimensions. 
Exploiting the third dimension offers new degrees of freedom leading to, for example, topological effects that arise from the curvature. 
Here, using numerical simulations, we uncover the low-temperature magnetic behaviour of a finite three-dimensional spin lattice: the buckyball artificial spin ice, where the spins are located on the edges of a regular buckyball.
This frustrated system has a non-trivial structural topology that results in a rich spectrum of thermal magnetic behaviour, beginning with a crossover from paramagnetism to a Spin-Ice sector followed by the formation of an imperfect charge crystal, before partial spin order is established in three separate steps. 
The final ground state configuration is described by a pair of robust topological magnetic defects that arise because of the finite curved nature of the spatial-spin interaction. 
Our work uncovers the intricate thermodynamics of the buckyball artificial spin ice. 
In doing so, we pave the way to designing unusual magnetic textures in other curved three-dimensional nanomagnetic systems by exploiting the interplay between structural topology and the dipolar interaction.
\end{abstract}
\begin{document}

\flushbottom
\maketitle

\thispagestyle{empty}

\section{\label{sec:Introduction}Introduction}
The paradigm that complex emergent behaviour can arise from combining simple building blocks~\cite{1972Anderson} is neatly epitomised by artificial spin ices.
These arrays of strongly-coupled nanomagnets~\cite{2020Skjaervo} can host complex collective behaviour including emergent magnetic monopoles~\cite{2010Ladak, 2011Mengotti}, vertex frustration~\cite{2016Gilbert} and entropy-driven phase transitions~\cite{2022Saglam}. 
Originally envisaged as an experimentally tractable mesoscopic analogue to the rare-earth pyrochlores~\cite{1997Harris}, artificial spin ices now offer a route to realising a plethora of model systems that address several important questions in statistical physics, and make it possible to explore a host of lattices, even going beyond those found in nature~\cite{2006Wang, 2016Perrin, 2017Gliga, 2024Yue}. 
By carefully tailoring the design of the lattice, and making use of modern lithographic techniques, different symmetries and spin dimensionalities can be realised in artificial spin ices~\cite{2018Leo, 2018Louis, 2023Zhang}, and their magnetic order can then be inspected in real space to observe thermal relaxation~\cite{2013Farhan} and phase transitions~\cite{2022Hofhuis} as they occur.

Until now, artificial spin ices have mainly been created in two dimensions (2D), since it is challenging to manufacture and characterise three-dimensional (3D) structures.
In order to observe their phase transitions, such 2D systems are made large enough to approach the bulk thermodynamic limit, avoiding finite size effects that smear out the appearance of critical phenomena~\cite{1976Landau}. 
Here we take a different approach and explicitly target the fascinating magnetic textures that can be hosted by a \emph{finite} spin lattice with curvature, namely one based on the buckyball.
The buckyball is a truncated icosahedron, a solid with both hexagonal and pentagonal faces, whose most widely-known realisation in condensed matter is the C$_{60}$ Buckminsterfullerene allotrope of carbon~\cite{1985Kroto}.

We construct our lattice by placing a point magnetic dipole on the midpoint of each edge of a regular buckyball, and term the resulting system the buckyball artificial spin ice. 
In our framework, the spins have an Ising degree of freedom and so are constrained to point in one of the two directions parallel to the orientation of their associated edge, i.e. into one vertex and out from the other.
Previously, the field-induced magnetisation reversal and the microwave response of such a connected lattice have been simulated using finite element techniques~\cite{2021Cheenikundil, 2022Cheenikundil}, showing that high-energy excitations can be seeded by an appropriately applied magnetic field.

Using Monte Carlo simulations, we reveal the thermodynamics of the buckyball artificial spin ice that, up until now, have remained completely unexplored.
We discover a  rich spectrum of thermal properties with five crossovers, which separate a wide hinterland of magnetic orders based on cooperative spin structures occurring at different length scales.
At high temperatures, the buckyball crosses over from paramagnetism into a Spin-Ice sector, which is characterised by short-range correlations and an ice rule governing the net number of spins that point into the vertices. 
On cooling the buckyball further, the net magnetic charges at these vertices order across the lattice.
However, perfect charge order cannot be accommodated because of the topology of buckyball, resulting in an imperfect charge crystal where the spins continue to fluctuate.
Remarkably, spin order is then established in three separate stages, with the ground state described by a pair of robust topological defects, whose formation can be controlled with an applied magnetic field. 

The richness of the behaviour arises because the spins form head-to-tail loops, which perceive the entire extent of the finite lattice.
What then makes these non-local loop structures---a hallmark of a system with ice-physics~\cite{2013Jaubert}---special is that they are embedded in a lattice with a non-trivial topology, which is effectively wrapped in on itself.
Here, long-range interactions between spins on opposite sides of the buckyball, mediated by these loops of head-to-tail spins, promote the unusual low-temperature crossovers.
Indeed, due to the curvature of the lattice, further-out neighbouring spins are closer than in a comparable, planar system.

The buckyball artificial spin ice is highly frustrated and many of the crossovers occur at temperature scales well-below the nearest-neighbour interaction.
This frustration arises through competing interactions on different length scales.
Firstly, there is a short-range frustration at the level of individual vertices, which exhibit a preference for an ice-rule configuration featuring one unsatisfied interaction.
Then, frustration arises at an intermediate length scale due to the impossibility of achieving a perfect arrangement of magnetic charges on the surface of the buckyball.
It is energetically favorable for every positive charge to be surrounded by nearby vertices with a negative charge and vice versa.
However, perfect charge order cannot be accommodated everywhere, leading to a frustration associated with their placement. 
Lastly, frustration arises on a larger length scale through the creation of topological defects. 
These defects can seed in various locations, giving rise to a configurational entropy.

This work highlights that finite 3D lattices can host unusual magnetic textures that have no obvious counterpart in naturally occurring microscopic materials. 
While we have limited ourselves to a computational investigation, we foresee that such 3D lattices based on Platonic and Archimedean solids could be routinely realised using existing lithographic techniques~\cite{2015Donnelly, 2019Gliga, 2017FernandezPacecho, 2019May, 2020Skoric, 2022Pip, 2024Volkov}.
However, magnetic imaging remains challenging, requiring further developments to capture the thermodynamics~\cite{2020Donnelly}. 

As we report, the buckyball artificial spin ice serves as a basis for a new form of ``mesoscale molecular magnet''~\cite{2020Coronado}.
In molecular magnets, magnetic moments arising from, say, metal ions or organic radicals, are embedded on specific sites within molecular compounds and can subsequently interact through \mbox{(super-)exchange coupling}~\cite{2008Chen}.
Through careful synthesis to control the placement of these moments, unusual phases can be realised including quantum spin liquids~\cite{2018Zhang}, with applications in organic spintronics~\cite{2008Bogani}. 
The buckyball artificial spin ice is a mesoscale analogue of these compounds where, as we reveal, exotic magnetic order arises through both local and global constraints imposed by the topology of the lattice.

\section{\label{sec:phase_diagram}Thermal Properties of the Buckyball Artificial Spin Ice}

\begin{figure*}
\centering
\includegraphics[width=1.0\textwidth]{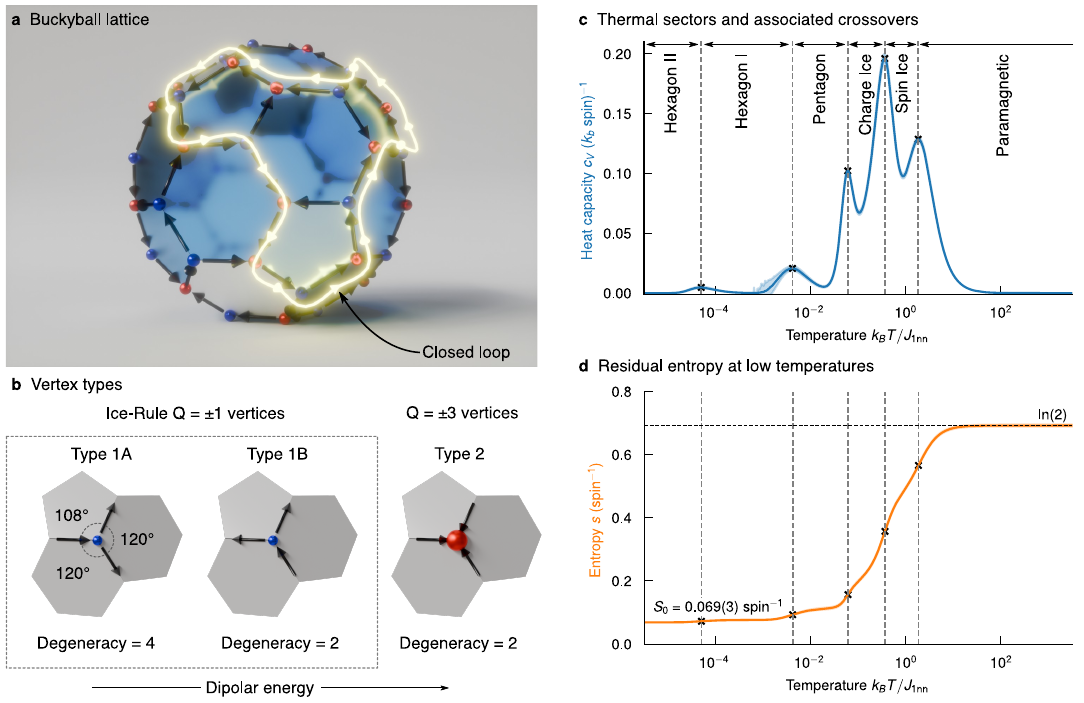}
\captionsetup{font=footnotesize}
\caption{\label{fig:main_lattice} \textbf{Thermal properties of the buckyball artificial spin ice.}
\textbf{a},~The buckyball artificial spin ice is constructed by placing point magnetic dipoles on the midpoints of the edges of a regular buckyball.
These spins have an Ising degree of freedom and are constrained to point in one of two directions along an edge. 
A closed loop of head-to-tail spins is highlighted. 
\textbf{b},~Example spin and vertex charge configurations for the three vertex types are shown, with the point dipolar energy increasing from left to right. 
Ice-rule $Q = \pm 1$ vertices have three spins that are either in a two-in/one-out or two-out/one-in configuration, but differ in energy depending on whether the two spins associated with a pentagonal face are in a favourable head-to-tail alignment (Type 1A configuration) or unfavourable alignment (Type 1B configuration).
In a high-energy $Q = \pm 3$ vertex, all three spins point either in or out.
We refer to these as Type~2 vertex configurations.
The degeneracy of each vertex configuration is indicated.
\textbf{c},\textbf{d},~Temperature dependence of \textbf{c},~the specific heat capacity $c_V$ of the buckyball artificial spin ice, which exhibits five peaks and \textbf{d},~the corresponding entropy~$s$.
The five peaks in the heat capacity are indicated with crosses and dashed vertical lines, and coincide with points of inflection in the entropy, also indicated with crosses.
Each of the peaks is associated with a crossover to a new type of magnetic order.
The names we give to each of the sectors are labelled explicitly in \textbf{c}.
The buckyball artificial spin ice still possesses a residual entropy per spin, $S_0 = 0.069(3)$, at $T = 0$ as indicated.
In panels \textbf{c} and \textbf{d}, shaded regions demarcate $\pm 1\sigma$ around the mean of $5\,000$ independent simulation runs.
}
\end{figure*}

The buckyball is one of the $13$ Archimedean solids, which are solids characterised by having faces that are two or more types of regular polygons.
The buckyball can be constructed from an icosahedron by truncating each of its vertices one third of the way along each edge.
For this reason, the buckyball is also called a \emph{truncated icosahedron}.
This results in both pentagonal and hexagonal faces of equal side lengths.
The buckyball has $60$ vertices, $90$ edges and $32$ faces, of which $12$ are pentagons and $20$ are hexagons.
Three edges meet at any vertex and the buckyball possesses vertex transitivity, so that all vertices are alike in terms of their local surroundings.
The edges of a buckyball can be sorted into two types: those that connect two hexagons, and those that connect a hexagon to a pentagon.

The buckyball artificial spin ice is constructed by placing a point magnetic dipole at the midpoint of each edge of a regular buckyball, as depicted in Fig.~\ref{fig:main_lattice}\textbf{a}.
We further constrain each point dipole to point in one of two directions along its associated edge.
In effect, each dipole represents the macrospin of an isolated single domain nanomagnet, which is the building block common to many experimental artificial spin ices~\cite{2020Skjaervo}.

The conventional 2D artificial spin ice geometry that most closely resembles our buckyball lattice is the artificial kagome spin ice.
Indeed, the surface of a buckyball bears similarities to a kagome lattice that has been wrapped around on itself to leave no boundaries. 
However, since this wrapping cannot be seamless, the buckyball includes pentagonal as well as hexagonal faces.
In this sense, we can view the buckyball lattice as like a kagome lattice in which an ordered arrangement of pentagonal defects has been introduced to accommodate the curvature~\cite{2002Nelson}.
The phase diagram of the in-plane artificial kagome spin ice~\cite{2009Moller, 2011Chern, 2012Chern} features one crossover from a high-temperature paramagnetic phase to a spin-ice phase.
This is followed by two phase transitions from the spin-ice phase to a charge-ordered phase and, finally, to a long-range spin-ordered state.
The ground state of the in-plane artificial kagome spin ice is six-fold degenerate, and is characterised by the head-to-tail alignment of spins associated with its hexagonal plaquettes---a configuration referred to as the `loop'~\cite{2009Moller} or `vortex'~\cite{2013Chopdekar} configuration.
In general, head-to-tail arrangements of magnetic moments minimise the stray magnetic field, so we would expect that the low-energy states of the buckyball artificial spin ice would feature them.
Indeed, one such closed loop is highlighted in Fig.~\ref{fig:main_lattice}\textbf{a}.

At each vertex in our buckyball lattice, we define a net magnetic charge, which is given by the number of spins that point into a vertex less the number that point out. 
Low energy vertices are those that obey an ice-rule: $Q = \pm 1$ with either a $2$-in-$1$-out or $2$-out-$1$-in spin configuration.
Unlike the artificial kagome spin ice, where all $Q = \pm 1$ vertices have the same energy, ice-rule vertices in the buckyball artificial spin ice can be split into two categories (see Fig.~\ref{fig:main_lattice}\textbf{b}).
This arises because the internal angles of a regular pentagon, $108^{\circ}$, are smaller than the internal angles of a regular hexagon, $120^{\circ}$.
As a consequence, at each vertex, the two spins that are associated with a pentagonal face are slightly closer together than either are to the remaining spin that bridges two hexagonal faces.
Of the six possible ice-rule configurations for a single vertex, four feature a favourable head-to-tail alignment of these pentagonal spins.
These Type~1A vertices are the lowest energy configurations for a single vertex.
The remaining two ice-rule vertex configurations, which we denote Type~1B, feature an unfavourable alignment of these pentagonal spins and are slightly higher in energy than the Type~1A vertices.
The highest energy vertices, Type~2, have charge $Q = \pm 3$ with all three spins pointing in or out (also shown in Fig.~\ref{fig:main_lattice}\textbf{b}).

To simulate the thermodynamic behaviour of the buckyball artificial spin ice, we perform extensive Monte Carlo simulations using the Metropolis-Hastings algorithm~\cite{1949Metropolis,1953Metropolis} incorporating both single spin flip dynamics and loop moves~\cite{1972Rahman, 1998Barkema, 2001Melko} on a point dipolar Hamiltonian.
Further details concerning the Monte Carlo simulations are given in the Methods section and in the Supplementary Information~\cite{SupplementalMaterial}.
In Fig.~\ref{fig:main_lattice}\textbf{c} and Fig.~\ref{fig:main_lattice}\textbf{d}, we show the behaviour of the specific heat capacity~$c_V$ (in blue) and the specific entropy~$s$ (in orange) as a function of the reduced temperature $k_B T/J_{1\mathrm{nn}}$, where $J_{1\mathrm{nn}}$ is the first-nearest-neighbour interaction strength and $k_B$ is the Boltzmann constant.
Surprisingly, the specific heat exhibits five peaks, with three at very low-temperatures well below the temperature scale associated with the nearest-neighbour interaction.
The highest temperature peak is relatively broad and occurs at ${k_B T/J_{1\mathrm{nn}} = 1.85}$, which is of the order of the first-nearest-neighbour interaction. 
Subsequent peaks, at ${k_B T/J_{1\mathrm{nn}} \approx 0.360, 0.061, 0.004}$~and~$0.0006$, become progressively sharper.
The origin of these peaks is not immediately obvious, since artificial spin ices to date have exhibited at most three peaks that are often associated with the formation of an ice-rule-obeying sector, a charge-ordered phase and a long-range spin-ordered state.
Moreover, it would be expected that finite-size effects would smear out the sharper peaks.
The appearance of so many peaks for the buckyball artificial spin ice suggests the formation of novel types of magnetic order, which are promoted by the topology of the lattice.

Each one of the peaks is associated with a point of inflection in the specific entropy $s$ as new configurations suddenly become accessible with thermal fluctuations at that temperature $T$.
Despite the inequivalent interactions at vertices, the buckyball artificial spin ice remains a frustrated system as a whole, with a residual entropy, $S_0 = 0.069(3)$ per spin.
Here, and in the Supplementary Information~\cite{SupplementalMaterial}, we use the notation that the number in parentheses indicates the $1\sigma$ uncertainty in the last digit of the averaged quantity, where $\sigma$ is the standard deviation.
By definition, the residual entropy is related to the natural logarithm of the number of ground states $\Omega$ through $S_0 = (1/N) \ln(\Omega)$, where $N = 90$ is the number of spins.
This suggests a ground state degeneracy of around $500$ magnetic configurations.

Interestingly, if we do not incorporate loop moves in our simulations, only four peaks are obtained in the specific heat capacity.
In this case, two of the low-temperature peaks are replaced by a single broad one at $k_B T/J_{1\mathrm{nn}}\approx 10^{-2}$ and the residual entropy is higher~(Supplementary Information, Section~I~\cite{SupplementalMaterial}), which is consistent with the notion that loops moves are necessary to access topological sectors beyond the reach of single spin flip dynamics~\cite{2013Jaubert}.
This effect is common to many ice models where the acceptance rate for single spin flips tends to zero before the onset of long-range order, and for which non-local dynamics must be explicitly introduced to maintain thermal equilibrium~\cite{2004Melko}.

In bulk systems, peaks in the specific heat reflect non-analytical behaviour in the free energy, and thus phase transition phenomena. 
In our finite system, the observed peaks indicate crossovers to different regimes of magnetic order, which will be here referred to as \emph{sectors}.
In the next section, we describe the spatial structure of the individual sectors that are labelled in Fig.~\ref{fig:main_lattice}\textbf{c}. 

\section{\label{sec:configurations}Real-space spin and charge configurations}

\begin{figure*}
\centering
\includegraphics[width=1.0\textwidth]{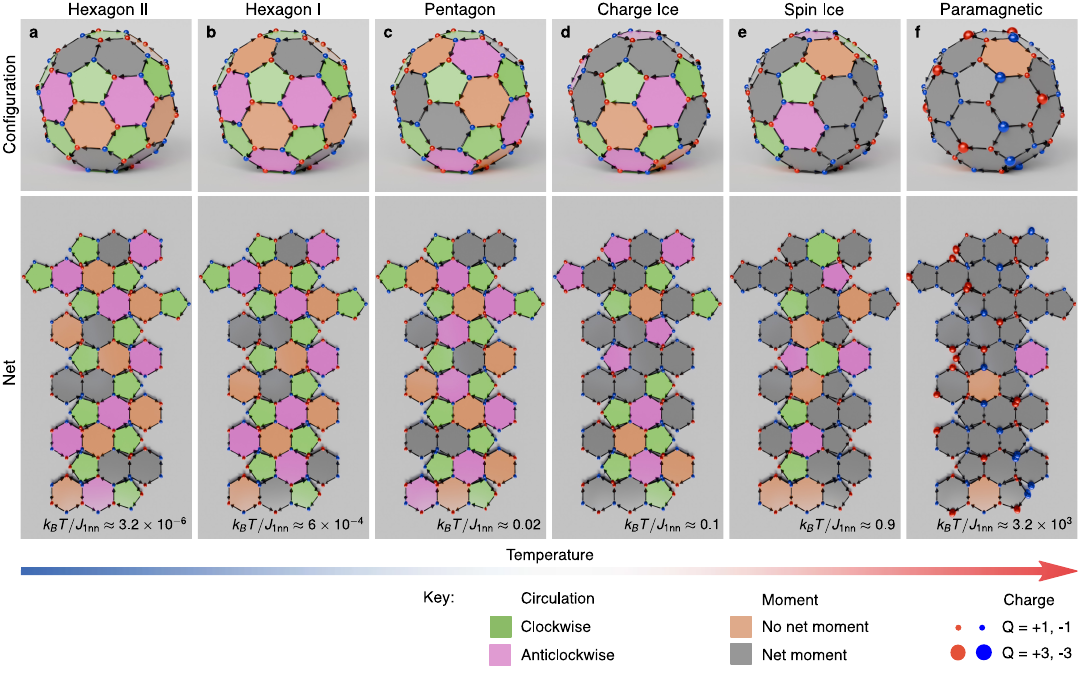}
\captionsetup{font=footnotesize}
\caption{\label{fig:main_configurations} \textbf{Example magnetic configurations taken from the six different sectors of the buckyball artificial spin ice.} 
Top panels: the 3D lattice, showing the direction of the spins (black arrows) and the charge at each vertex. 
Positive (negative) charges are shown in red (blue) with the sizes indicating whether the charges are $\pm 1$~(small) or $\pm 3$~(large).
Bottom panels: net representations of the magnetic configuration. 
For both the 3D lattices and the net representations, the polygonal faces are coloured according to the configuration of their associated spins: green (pink) for a clockwise (anticlockwise) loop of spins, grey if the face has a net moment, and orange if the face has neither a net moment nor a closed loop.
The sense of circulation is defined as that on viewing the buckyball from the outside along the normal to the given face.
The reduced temperature is given in the bottom right of each net and increases from left to right.
}
\end{figure*}
In Fig.~\ref{fig:main_configurations}, we show typical spin and charge configurations taken from each of the six different sectors of the buckyball artificial spin ice, with the temperature increasing from left to right as indicated. 
In addition to the 3D representation, we reveal the full configuration with a net representation, where the buckyball is unfolded into a single planar structure~\cite{2018Friedman}. 
It should be noted that, while the number of faces in the net is the same as the number of faces in the solid, several vertices in the net may correspond to the same vertex in the solid, and similarly for edges.
Each configuration is taken from the middle of the temperature range associated with that particular sector in order to avoid fluctuations in the energy (and, hence, the spin configuration) that are more common at the crossovers.
This is because the order associated with a certain sector may not yet be fully apparent in the vicinity of the crossover to that sector, just as with a phase transition, where many differently sized clusters of the new phase exist close to the critical temperature.

At high temperatures (Fig.~\ref{fig:main_configurations}\textbf{f}), the buckyball artificial spin ice is in a paramagnetic state, displaying neither spin nor charge order with both ice-rule-obeying and high-energy $Q = \pm 3$ vertices present.
Most faces---whether hexagonal or pentagonal---have a net moment, as expected for a state with randomly-oriented spins.
As the buckyball artificial spin ice is cooled below $k_B T/J_{1\mathrm{nn}} \approx 1.85$, it enters a Spin-Ice sector~(Fig.~\ref{fig:main_configurations}\textbf{e}), where only $Q = \pm 1$ vertices are present.
At $k_B T/J_{1\mathrm{nn}} \approx 0.360$, the charges on the vertices of the buckyball begin to order and each $Q = +1$ vertex tries to surround itself with three $Q = -1$ vertices, and vice versa.
We therefore refer to this as the Charge-Ice sector.
On closer inspection of the configuration in Fig.~\ref{fig:main_configurations}\textbf{d}, we see that the spins associated with each of the pentagons have aligned head-to-tail in a loop, although not all loops are necessarily in the same sense with the clockwise (anticlockwise) pentagons shown in green (pink).
For the naming of these first three sectors, we have drawn a deliberate parallel with the phases of the artificial kagome spin ice, although we will later make clear that there are significant differences. 
As the temperature continues to decrease, we see that, at $k_B T/J_{1\mathrm{nn}} = 0.004$, there is the first emergence of a type of spin order that spans the entire lattice: the spins around the pentagonal faces now have the same sense of circulation and, accordingly, they are all coloured green in Fig.~\ref{fig:main_configurations}\textbf{c}. 
We therefore label this the Pentagon sector.
Some of the hexagons also have closed loops of spins but the ordering in terms of their sense of circulation is not correlated.
Since such closed loops are naturally the arrangement that minimises the stray field, we will refer to them as `flux-closed' faces.
The remaining two configurations~(Fig.~\ref{fig:main_configurations}\textbf{b}~and~Fig.~\ref{fig:main_configurations}\textbf{a}) are taken from the two low-temperature sectors.
These two sectors differ in how the hexagonal faces with a net moment are distributed across the surface of the solid, and we have thus named them Hexagon~I and Hexagon~II.
We unravel what these two crossovers signify in greater detail later in the paper but, suffice to say, they stem from different groups of spins freezing at different temperatures in the buckyball artificial spin ice, leading to the formation of different low-energy motifs that together tile the lattice.

\section{\label{sec:correlations}Role of long-range correlations in promoting the low-temperature crossovers}

\begin{figure*}
\centering
\includegraphics[width=0.75\textwidth]{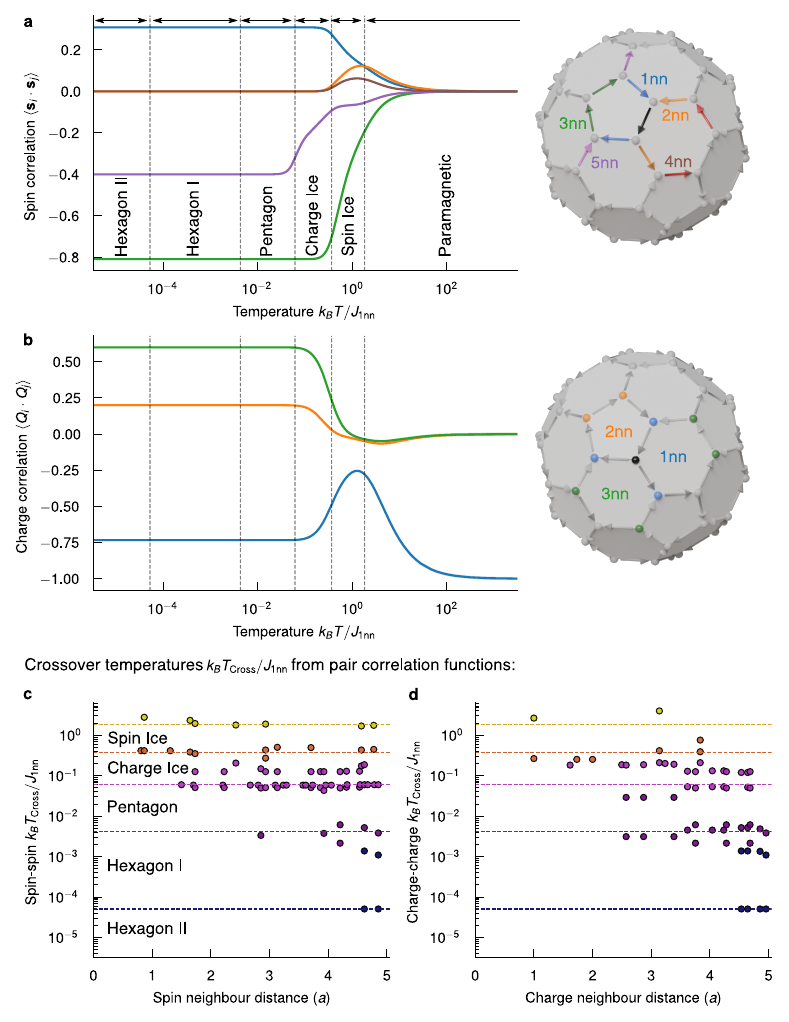}
\captionsetup{font=footnotesize}
\caption{\label{fig:main_correlations} \textbf{Spin and charge correlations in the buckyball artificial spin ice.} 
\textbf{a},\textbf{b},~Temperature dependence of the first five nearest-neighbour spin correlations and first three nearest-neighbour charge correlations.
In each case, the definitions of the spin or charge neighbours, with respect to the central black spin or vertex, are given to the right.
The vertical dashed lines indicate the locations of the five crossover temperatures for the buckyball artificial spin ice.
In terms of the spin correlations, it is the behaviour of the first and second nearest neighbours that enforces the existence of an ice rule at each vertex.
As a result of the topology of the buckyball, perfect charge order cannot be accommodated, so that the value for $\langle Q_i Q_j \rangle_{\mathrm{1nn}}$ pairs (blue curve, panel~\textbf{b}) reaches a low temperature value of $-11/15$, as opposed to~$-1$ for the artificial kagome spin ice~\cite{2020Hofhuis}. 
Error bars are smaller than the line width.
\textbf{c},\textbf{d},~A crossover temperature $T_{\mathrm{Cross}}^{(n)}$ can be assigned to each spin or charge pair by finding the temperature at which the corresponding correlation function has a point of inflection (as described in the Methods).
These crossover temperatures are shown as a function of the spin and charge neighbour distance in \textbf{c} and \textbf{d}, respectively.
Each individual crossover temperature is coloured according to the nearest system-wide crossover, matching the colour of the relevant horizontal dashed line.
For the most part, the pairwise crossover temperatures lie close to one of the five system-wide crossover temperatures allowing us to determine which interactions are responsible for the emergence of each sector.
It is apparent from these plots that long-range interactions are responsible for the low temperature peaks, since data points above a neighbour separation of $2.5 a$ tend to cluster near to system-wide crossovers to the Pentagon, Hexagon~I and Hexagon~II sectors.
}
\end{figure*}

To provide an explanation for the existence of the different magnetic sectors, and the crossovers that delineate them, we consider the pairwise spin and charge correlations in the buckyball artificial spin ice. 
This is particularly important for the low-temperature crossovers because, from the combined spin and vertex maps alone, it is hard to say what defines exactly the Pentagon, Hexagon~I and Hexagon~II sectors.
Similarly, the low-temperature crossovers are are also not associated with any change in the loop structure on the buckyball or in the number of closed-loop faces~(Supplementary Information, Sections~I and~V~\cite{SupplementalMaterial}).
While the spatial structure of the Spin-Ice sector and the Charge-Ice sector are easier to grasp from the real-space configurations, their formation in a finite lattice is still surprising.
It is only by considering the spin and charge correlations that we gain a complete understanding of which interactions play a determining role in establishing the different magnetic sectors. 

The spin correlations $\langle \mathbf{s}_i \cdot \mathbf{s}_j \rangle$, up to the fifth-nearest-neighbour spin pair, and the charge correlations $\langle Q_i Q_j \rangle$, up to the third-nearest-neighbour charge pair, are displayed in Fig.~\ref{fig:main_correlations}\textbf{a} and Fig.~\ref{fig:main_correlations}\textbf{b}, respectively. 
The definitions of these nearest-neighbour pairs are given by the associated schematics.
The dashed vertical lines indicate the locations of the five peaks observed in the specific heat capacity of the buckyball artificial spin ice shown in Fig.~\ref{fig:main_lattice}\textbf{c}.

From the behaviour of these correlations, we note the following three key points: 
\begin{enumerate}
    \item The interactions between first- and second-nearest-neighbour spins, which completely encompass all interactions at a single vertex, are responsible for the emergence of the Spin-Ice sector. 
    Here the corresponding spin-spin correlators have maxima or inflection points at, or near to, the crossover to the Spin-Ice sector, as shown in Fig.~\ref{fig:main_correlations}\textbf{a}. 
    Signatures of this crossover to the Spin-Ice sector can also be seen in the other nearby correlation functions, in particular those for $4$nn and $5$nn. 
    This reflects the fact that there needs to be some cooperation between vertices to enforce an ice-rule everywhere, since setting a two-out/one-in or two-in/one-out configuration on one vertex necessarily constrains the possible ice-rule configurations on adjacent vertices. 
    In fact, the situation is slightly more nuanced due to the `broken' nearest-neighbour interactions at a vertex, which split the energy of ice-rule states.
    Evidence of this is found in the temperature dependence of the three vertex populations, Type~1A, Type~1B and high energy $Q = \pm3$ Type~2, which are shown in the Supplementary Information, Section~IV~\cite{SupplementalMaterial}.
    There it can be seen that, at the crossover to the Spin-Ice sector, high-energy vertices are no longer present and all vertices obey the ice rule. 
    At the crossover to the Charge-Ice sector, the higher-energy Type~1B vertices also disappear, and only the four degenerate lowest-energy Type~1A vertices are found.
    
    \item As a result of the topology of the buckyball, perfect charge order cannot be accommodated.
    In the Charge-Ice sector, some $Q = +1$ vertices exist that are not surrounded by three $Q = -1$ vertices, and vice versa.
    As a result, the value for $\langle Q_i Q_j \rangle_{\mathrm{1nn}}$ pairs reaches a low-temperature value of only $-11/15$, as opposed to~$-1$ for the artificial kagome spin ice. 
    This highlights how the buckyball artificial spin ice is an extremely frustrated system, since not only is one of the interactions at each ice-rule vertex unsatisfied. 
    Rather, the topology of the buckyball means that the vertices themselves cannot be split into bipartite sets.
    There is then a frustration that arises on a larger length scale, namely between vertices, as a consequence of the imperfect arrangement of the charges. 

    \item For the subset of spin and charge correlations that we choose to display in Fig.~\ref{fig:main_correlations}\textbf{a},\textbf{b}, there is little change in the correlation functions in the vicinity of the crossovers to the Pentagon, Hexagon~I, or Hexagon~II sectors. 
    This suggests that correlations between pairs of spins and charges that are further away from each other must be responsible for the existence of these low-temperature peaks in the heat capacity (Fig.~\ref{fig:main_lattice}\textbf{c}).  
\end{enumerate}

To clarify the third observation, we assign individual \emph{crossover} temperatures to each of the nearest-neighbour spin and charge correlation pairs, by finding the temperatures at which which the corresponding correlation functions have a point of inflection (see Methods for more details).
For the case of two spins, which are $n$th-nearest neighbours separated by a distance $r_n$, we can understand a crossover temperature $T_{\mathrm{Cross}}^{(n)}$ as the temperature at which fluctuations between $n$th-nearest neighbours are at a local extremum.
As the temperature is lowered, the correlations change in size until, at their lowest crossover temperature, the interaction freezes out completely.
At this point, the temperature scale is such that this interaction plays no further role in the ordering. 
A similar logic applies to a pair of neighbouring charges.

The crossover temperatures for these individual spin and charge correlators, plotted as a function of the neighbour distance, are displayed in Fig.~\ref{fig:main_correlations}\textbf{c} and Fig.~\ref{fig:main_correlations}\textbf{d}, respectively.
Here, the horizontal dashed lines indicate the five crossover temperatures of the buckyball artificial spin ice as a whole.
The majority of the lowest spin-spin crossover temperatures lie close to the crossover to the Pentagon sector, which is the first temperature at which some form of system-spanning spin order---with all pentagons displaying loops of spins in the same sense---is established. 
However, the correlations between a number of furthest-neighbour-spin pairs still vary with temperature in the vicinity of the two lowest-temperature crossovers.
In terms of the charge-charge crossover temperatures, many of these lie near to the transition to the Charge-Ice sector, as expected since this is the point at which charge order is established on the buckyball surface.
Beyond a separation of $2.5a$, many charge-charge interactions have a lowest $T_{\mathrm{Cross}}^{(n)}$ close to the Hexagon~I or even Hexagon~II transition. 
We have thus established that, rather than the nearest-neighbour interactions, it is the long-range interactions that are responsible for the low-temperature crossovers.
This is not the complete picture because, as we now demonstrate, the low-temperature magnetic structure is additionally constrained by the geometry of our curved finite system, which means that further-out neighbours are closer than they would be in a 2D artificial spin ice.

\section{\label{sec:three_stage_spin_ordering} Three-stage spin ordering and ground state motifs}
\begin{figure*}
\centering
\includegraphics[width=\textwidth]{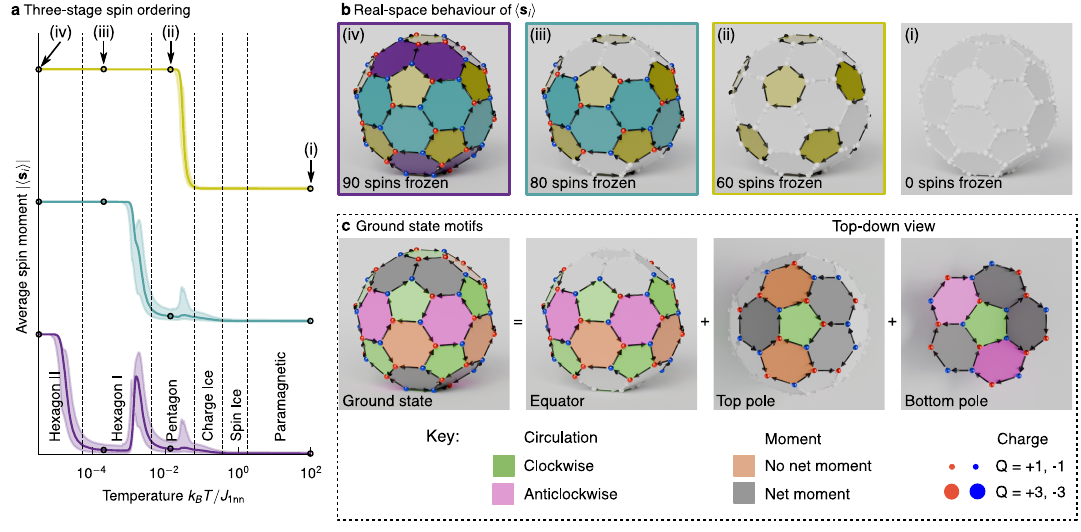}
\captionsetup{font=footnotesize}
\caption{\label{fig:main_spin_ordering} \textbf{Three-stage spin ordering in the buckyball artificial spin ice.}
\textbf{a}, The temperature dependence of the absolute value of the average spin moment $\langle {\textbf{s}}_i \rangle$.
The 90 spins in the buckyball are sorted into three groups, which order separately in the Pentagon, Hexagon~I, and Hexagon~II sectors, and are shown by the three curves in gold, turquoise and indigo, with each curve displaced in the $y$-direction.
Each curve begins at $0$ in the high-temperature limit and reaches $1$ by the lowest temperature simulated. 
For each group, shaded regions indicate the $95\%$ confidence interval around the mean of $5\,000$~independent runs. 
Uncertainties are calculated separately for data above and below the mean, giving asymmetric error bars, which ensures that the mean and its uncertainty does not go above one.
\textbf{b}, Real-space representations of the average spin $\langle {\textbf{s}}_i \rangle$, corresponding to the temperatures marked in panel \textbf{a} with labels and circular markers.
The faces of the buckyball are coloured according to the temperature at which all their associated spins freeze, matching the colour palette of panel \textbf{a}.
The spins appear in black only once they have frozen; otherwise, they appear as white, double-headed arrows to indicate that they continue to fluctuate.
The states are: (i)~at high temperature, when no spins have ordered; (ii)~just below the crossover to the Pentagon sector, when the $60$ spins associated with the pentagonal faces have ordered (gold); (iii)~below the crossover to the Hexagon~I sector, when a further $20$ spins have ordered so that a belt of hexagons is now frozen (turquoise); and (iv)~below the crossover to the  Hexagon~II sector, when the remaining $10$~spins have ordered, so that the spins associated with $5$~hexagonal faces either side of the belt freeze (indigo).
\textbf{c},~The ground state of the buckyball artificial spin ice can be decomposed into three motifs: an equatorial belt, formed from the maximal number of possible closed-loop pentagons and hexagons; and two poles, top and bottom, which are formed around a central closed-loop pentagon.
In this panel, the faces are coloured according to the circulation of their associated spins as given by the key, which matches that used in Fig.~\ref{fig:main_configurations}.
}
\end{figure*}

We now show precisely how spin order is progressively established in each of the Pentagon, Hexagon~I and Hexagon~II sectors. 
For this, in analogy with the total magnetisation of a magnetic solid, we consider the temperature dependence of $\langle \textbf{s}_i \rangle$, whose absolute value encodes the average length of the spin vector $\textbf{s}_i$ averaged over the number of Monte Carlo steps performed at that temperature $T$.
Then, spin $i$ can be said to be frozen when $\langle \textbf{s}_i \rangle = \pm1$.

On plotting the absolute value of this quantity for each spin in the buckyball artificial spin ice in Fig.~\ref{fig:main_spin_ordering}\textbf{a}, we observe that the spins fall into three groups (curves in gold, turquoise, and and indigo), which freeze at different temperatures near to the low-temperature crossovers.
To illustrate better what these groups of spins represent, we show the real-space configurations of $\langle \textbf{s}_i \rangle$ in Fig.~\ref{fig:main_spin_ordering}\textbf{b} corresponding to the four temperatures indicated with black arrows in panel Fig.~\ref{fig:main_spin_ordering}\textbf{a}.
Here, the spins appear in black only when they have frozen; otherwise, they are drawn as white double-headed arrows indicating that they continue to fluctuate.
The faces of the buckyball artificial spin ice are coloured accorded to the lowest ordering temperature of their associated spins, in a colour scheme consistent with the three groups of spins in Fig.~\ref{fig:main_spin_ordering}\textbf{a}. 
The highlighted states are:
\begin{enumerate}[label=(\roman*)]
    \item at high temperature, when no spins have ordered so that none of the faces are coloured.
    \item below the crossover to the Pentagon sector, when the $60$~spins associated with all $12$ pentagons freeze in closed loops with the same sense of circulation (faces coloured gold).
    \item below the crossover to the Hexagon~I sector, when a further $20$~spins freeze.
    These spins are associated with $10$~hexagonal faces, shown in turquoise, which form a chain round the equator of the buckyball.
    \item below the crossover to the Hexagon~II sector, when the remaining $10$~spins freeze.
    At this point, the final ten hexagonal faces---five on either side of the buckyball at the poles---now order, as shaded in indigo.
\end{enumerate}

We now have all the information we need to completely describe the spin motifs that together make up the ground state of the buckyball artificial spin ice (Fig.~\ref{fig:main_spin_ordering}\textbf{c}). 
This ground state is formed from a belt of spins around the equator, which features the maximal number of closed-loop faces.
In this belt, the loops associated with the pentagons and the hexagons have opposite sense; in this example, clockwise and anticlockwise (in green and pink), respectively.
It is not possible that every hexagon in this belt flux-closes on account of the topology. 
Hence, there are some hexagonal faces that are not flux-closed but which have no overall net moment (shaded in orange).

This leaves $10$ hexagonal faces, five on either side of the buckyball.
Each set of five hexagonal faces surround a central flux-closed pentagon.
These two `poles' of the buckyball artificial spin ice involve some hexagonal faces with a net moment, which are coloured grey.
Thus, the spin configuration at these poles is locally higher in energy than an arrangement in which the spins associated with the majority of the faces are flux-closed, as occurs around the equator.
However, complete flux-closure cannot be accommodated everywhere.
These poles are then examples of topological defects in the truest sense: they are constrained by the geometry of the lattice and must form as a consequence of the low-energy flux-closed motif around the equator but, crucially, they cannot be eliminated by a smooth deformation or flips of nanomagnet orientations~\cite{1995Chaikin}.
These constitute the most intriguing instances of topological defects in an artificial spin ice context, since previous examples have relied on physically distorting the lattice at certain positions to promote their formation~\cite{2017Drisko, 2022Puttock}.
Indeed, in the buckyball artificial spin ice, these topological defects can form in any of six different locations around pairs of diametrically-opposite pentagons.
By symmetry, it is possible to show that there are $480$ permutations of such defects pairs, which results in a residual entropy of $S_0 \approx 0.0685$ per spin.
This agrees extremely well with the value obtained through numerical integration of the specific heat capacity, $S_0 \approx 0.069(3)$ per spin, given in Fig.~\ref{fig:main_lattice}\textbf{d}.

\section{\label{sec:discussion}Conclusion}

Our buckyball artificial spin ice exhibits a diverse spectrum of thermal behaviour, characterised by five crossovers that separate six different magnetic sectors, beginning at higher temperatures with a crossover from paramagnetic behaviour to a Spin-Ice sector and then to an imperfect charge-ordered crystal.
Remarkably, we find that spin order is established via a three-stage mechanism, which is mediated by dipolar interactions between spins on opposite sides of the buckyball. 
The ground state itself hosts a pair of robust topological defect `poles' that, although locally high in energy, must be present so that flux-closed motifs can form elsewhere.
As with the vertex and charge frustration, these defects carry a certain entropy on account of their multiple configurations.
In addition, the existence of these topological defects means that complete compensation of the magnetic moment is not possible.
Hence, the buckyball artificial spin ice possesses a small net moment even in the ground state, which is orientated approximately along the normal to the pentagonal face at the centre of each pole.
In terms of applications, one could then envisage using these 3D finite lattices---or, indeed, assemblies of them---as ultra-sensitive 3D magnetic field sensors, with the location of the topological defects depending on the orientation of the applied magnetic field.

We have presented the first characterisation of the magnetic ordering that occurs in a fascinating, albeit specific, example of an Archimedean solid. 
We envision that further complex thermal behaviour and new types of magnetic order await to be found in different members of the Platonic, Archimedean, and Catalan solid families, since these lattices all involve polygons with different vertex coordinations and effective local curvature. 

In effect, these finite lattices present a new form of artificial molecular magnet, where the moments are situated on the edges rather than the vertices of the solid.
Indeed, in a buckyball lattice with the spins located on the vertices as would be found in molecules, the thermal behaviour is much simpler with only a single crossover (see Supplementary Information, Section VI~\cite{SupplementalMaterial}).
In contrast, the richness of the physics we observe in our buckyball artificial spin ice stems from the formation of loops of spins linking ice-rule vertices, which are embedded on a lattice with non-trivial topology.

Now that we have a complete understanding of the buckyball, we can extend our analysis to the the general family of fullerenes, of which the buckyball is one of the simplest examples.
This would offer the opportunity to study the physics of mobile topological defects. 
In the buckyball artificial spin ice, the topological defects are confined: they form but are not able to move even in the presence of thermal fluctuations, on account of their large size with respect to the finite size of the solid, since each topological defect involves the spins lining six faces, out of a total of 32 faces. 
For larger fullerenes, we would still expect topological defects to form initially around the pentagons to accommodate the curvature, but each individual defect may be less strongly coupled to any of its neighbours.
As a result, the defects should be able to depin from the pentagons and move across the lattice at finite temperature.
As we continue to increase the size of the fullerenes, we would expect the specific residual entropy to tend to zero, since the limiting case of such lattices is the artificial kagome spin ice lattice, which does not have an extensive residual entropy~\cite{2009Moller}. 
Seen in this context, we can understand the finite residual entropy in our buckyball artificial spin ice to be an expression of the various configurations that each topological defect can take. 

Our findings illuminate the rich landscape of magnetic phenomena that can emerge from the interplay of curvature, constraints and interactions in finite three-dimensional `nanomagnetic molecules'. 
While the fabrication of such lattices using current lithographic techniques is feasible~\cite{2015Donnelly, 2019Gliga, 2020Ren}, probing their thermodynamics remains a significant challenge, with only the frozen states at surfaces imaged with magnetic force microscopy~\cite{2019May, 2021May, 2023Saccone} and the magnetic configuration of small structures probed with synchrotron x-ray techniques up to now~\cite{2022Pip}. 
Overcoming this hurdle will require innovative approaches in terms of fabrication and further development of advanced imaging methodologies~\cite{2020Donnelly}.

\section*{Acknowledgements}
This work was carried out under the auspices of the Swiss National Science Foundation (project no. 200020\_200332).
Part of this work was performed on the Merlin6 High Performance Computing Cluster at the Paul Scherrer Institute, Villigen.
G.M.M. thanks Hugo Bocquet for helpful discussions regarding ways to classify the symmetry of the magnetic ground state, and Martin Macauley for useful comments on drafts of this work.

\section*{Author contributions}
L.J.H. first motivated this study, and provided supervision.
G.M.M. wrote the Monte Carlo code and, with the help of P.M.D. and L.B., analysed the results.
G.M.M. and L.J.H wrote the manuscript,  with the input of the other authors.

\section*{Competing interests}
The authors declare no competing interests. 

\section*{Methods}

\subsection*{Monte Carlo Simulations}

The dipolar Hamiltonian of the buckyball artificial spin ice is
\begin{linenomath*}
\begin{equation}
\centering
\label{main:eq:dipolar_hamiltonian}
\mathcal{H} = D \sum_{i \neq j}^{N} \left[ \frac{\hat{\mathbf{s}}_i \cdot \hat{\mathbf{s}}_j}{r_{ij}^3} - \frac{3(\hat{\mathbf{s}}_i \cdot \mathbf{r}_{ij}) (\hat{\mathbf{s}}_j \cdot \mathbf{r}_{ij})}{r_{ij}^5}\right],
\end{equation}
\end{linenomath*}
such that spin $i$, located at position $\mathbf{r}_i$, has a normalised moment $\hat{\mathbf{s}}_i$.
The vector $\mathbf{r}_{ij} \equiv  \mathbf{r}_j - \mathbf{r}_i $ points from spin $i$ to spin $j$. 
The energy scale of the interaction is set through the dipolar constant, $D = \mu_0 (M_S V)^2/(4\pi a^3)$, where $M_S$ and $V$ are the saturation magnetisation and volume of each nanomagnet, respectively.
The symbol $\mu_0$ denotes the magnetic permeability of free space.
The parameter $a$, which is the lattice constant, controls the scale of the distances between spins.
Here, we set $a = 1$ since the spin and vertex positions have been calculated for a buckyball with unit edge length.
For convenience, we work in reduced units, and re-scale temperatures in terms of the largest dipolar energy, which is the first-nearest-neighbour interaction strength $J_{1\mathrm{nn}} \approx 3.1424~D$.

The Monte Carlo simulations were performed using both single spin flips and loop moves. 
In this context, a loop is any closed head-to-tail chain of spins.
When spins in a loop are flipped simultaneously, the charge at each vertex remains the same. 
This is because every loop must enter and exit a vertex, which ensures local conservation of charge. 
Once such loop moves are proposed, they are accepted or rejected according to the Metropolis criterion, just as with single spin flips. 
This criterion dictates that the proposed move is accepted if it lowers the energy of the system outright.
Otherwise, it is accepted with a probability proportional to the Boltzmann factor $\exp(-\Delta E/ k_bT)$, where $\Delta E$ is the energy difference between the initial state and the proposed state.
Loop moves help to efficiently sample the low-energy states of a frustrated system~\cite{1977Hohenberg, 2018Hamp}, when either the waiting time to access the new state is prohibitively long, or where single spin flips are frozen out completely.
This often happens when, for example, constraints arising from the topology of the lattice promote the formation of motifs that incorporate a number of spins.
Loop moves then allow the system to move between these different `topological sectors', such as in the Coulomb phase of real spin ice , whose hallmark is algebraic spin-spin correlations~\cite{2013Jaubert, 1977Hohenberg, 2018Hamp}.

The specific heat capacity is calculated by recording the fluctuations in energy about the mean value $\langle E \rangle$ through
\begin{linenomath*}
\begin{equation}
\label{eq:specific_heat}
{\centering
c_V(T) = \frac{1}{N}\frac{\langle (E - \langle E \rangle ) ^ 2 \rangle}{k_BT^2},}
\end{equation}
\end{linenomath*}
\noindent where $N = 90$ is the number of spins in the buckyball, which has total energy $E$ at temperature $T$.
The entropy per spin at temperature~$T$ is obtained by numerical integration according to
\begin{linenomath*}
\begin{equation}
\label{eq:delta_S}
\centering
s(T) = \ln(2) - \int_{\infty}^T \frac{c_V}{T'} \mathrm{d}T'.
\end{equation}
\end{linenomath*}
The first term on the right hand side, $\ln(2)$, is the appropriate entropy in the high-temperature limit for a spin with an Ising degree of freedom, i.e. with two equally-likely states.

Verification of the Monte Carlo simulations is given in the Supplementary Information~\cite{SupplementalMaterial}.
This includes: (i)~tests on the robustness of the equilibration; (ii)~proof of the necessity to include loop moves to recover the full spectrum of thermal behaviour; and (iii)~determination of the required frequency of such moves to the reach the ground state.
For the simulation results presented in the main text, the full dipolar sum given in Eq.~(\ref{main:eq:dipolar_hamiltonian}) is implemented.
In Section~II in the Supplementary Information, we discuss the effect of implementing a cutoff radius beyond which interactions are neglected~\cite{SupplementalMaterial}.
As for the spin and charge correlations, truncating the dipolar interaction  demonstrates the important role that long-range interactions in play in the low-temperature crossovers in the buckyball artificial spin ice. 
The five peaks in the heat capacity are only recovered when \emph{all} interactions are included because of the 3D nature of the lattice.

\subsection*{Assigning crossover temperatures}
In Section~\ref{sec:correlations}, we discussed the allocation of crossover temperatures to individual spin and charge correlator pairs.
This approach is analogous to the way in which phase transitions correspond to changes in the curvature of the internal energy $U$ that, in turn, can appear as local maxima in $C_V = \mathrm{d}U/\mathrm{d}T$. 
For the spin-spin correlation between $n$th-nearest neighbours, $\langle \mathbf{s}_i \cdot \mathbf{s}_j \rangle^{(n)}$, we define the crossover temperature $T_{\mathrm{Cross}}^{(n)}$  as the temperature for which the quantity,
\begin{linenomath*}
\begin{equation}
\label{eq:crossoverT}
\centering
\frac{\mathrm{d}}{\mathrm{d}T} \langle \mathbf{s}_i \cdot \mathbf{s}_j \rangle^{(n)},
\end{equation}
\end{linenomath*}
has a local maximum \emph{or} minimum. 
In practise, we do this through numerically solving 
\begin{linenomath*}
\begin{equation}
\label{eq:second_deriv}
\centering
\frac{\mathrm{d}^2}{\mathrm{d}T^2} \langle \mathbf{s}_i \cdot \mathbf{s}_j \rangle^{(n)} = 0
\end{equation}
\end{linenomath*}
for $T$, and checking that the second derivative changes sign around that point. In general, a given correlation function may have more than one point of inflection. 

\section*{Supplementary Information}
See Supplementary Information at \hl{http://000.000.000.000} for further details on the Monte Carlo simulations; the effect of including a cutoff in the dipolar interaction; an alternative `dipolar needle' approximation for the interactions in the buckyball artificial spin ice; the temperature dependence of vertex populations; statistics about the loop population in each sector; and an alternative model, in which spins are placed on the vertices of the lattice~\cite{SupplementalMaterial}.

\section*{Data and code availability}
The data underpinning the results in this paper are available at the Zenodo repository: \hl{http://000.000.000.000}~\cite{Zenodo}.
The Monte Carlo code, written in Julia, is available on an online repository~\cite{Github}.
Further details can be requested from gavin.macauley@psi.ch.

\newpage

\providecommand{\noopsort}[1]{}\providecommand{\singleletter}[1]{#1}%

\end{document}